\documentclass[secnumarabic, pra, showkeys, showpacs]{revtex4}
\usepackage{bm}
\usepackage{graphicx}

\begin{document}
\title{Experiments performed in order to reveal fundamental differences between the diffraction and interference of waves and electrons}
\author{Victor V. Demjanov}
\affiliation{Ushakov Maritime State Academy, Novorossyisk, Russia}
\email{demjanov@nsma.ru}
\date{20 February, 2010}

\begin{abstract}
Diffraction patterns of electrons are believed to resemble those of electromagnetic waves (EMW). I performed a series of experiments invoked to show that the periodicity of peaks in the diffraction diagram of electrons is concerned with the periodicity of the arrangement of scattering centers in the diffraction grating in combination with the supposed character of the spatial structure of the electron as a system of regularly spaced concentric shells of elasticity. I started from the experiment on the diffraction of electrons and EMWs at the sharp edge of the opaque half-plane. This simple scattering configuration enabled me to discriminate between the re-radiation mechanism of the wave diffraction and ricochet scattering of electrons on the edge of the half-plane. Then I made experiments with scattering on composite objects proceeding step by step from a single straight edge to a couple of edges (one slit) and then to four edges (two slits). Thus I succeeded in interpretation of the double-slit diffraction (four straight edges) in terms of the scattering on a single edge.
In favor of the elastic corpuscle mechanism of the electron's scattering there speak also following experiments made by me. The displacement of a one half-plane of the slit along the direction of the electron beam at a wave half-length does not virtually affect the scattering diagram of electrons while the interference pattern of EMWs changes in this case from the positive to negative. On the other hand, the heating of the slits blurs and widens the scattering pattern of electrons but exerts no appreciable influence on that of EMWs.

I succeeded also in the observation of the electron's trajectory. Equipping the slit's edge with a semiconductor sensors I registered the passing of the electron through the slit detecting simultaneously a weaker signal at another slit. This technique of observation of the event what slit did the electron pass through appeared not to destroy or disturb the "interference" pattern.

Closing one slit by inserting therein a transparent for EMWs piece of dielectric I have found that the scattering pattern beneath this slit vanishes while the scattering pattern under the open slit remains as before when both slits were open. This experiment indicates that the electron "interferes" with its electromagnetic component passed via the slit plugged by the transparent for it dielectric.
\end{abstract}
\maketitle
\section{The aim of the research}

Since the experiments of G.P.Thomson \cite{Thomson} and C.Davisson$\&$L.H. Germer \cite{Davisson}  there established the general conviction that particles in many respects diffract likewise electromagnetic waves (EMW). The properties of waves include among others the re-radiation on scattering centers and interference. The features peculiar to corpuscles are the recoil from scattering centers and motion along trajectories. The interference pattern of waves looks as a series of lateral peaks. The similar form has the distribution of electrons after their scattering by a crystal lattice \cite{Thomson,Davisson} . The question under suspicion is if the regularity of the electron's scattering pattern may be in a sense a replication of the periodic structure of the crystal. In order to verify this supposition I performed scattering experiments, both for EMWs and electrons, on simple configurations which are protomorphic to the diffraction grating. Besides I equipped scattering objects with detectors which enabled me to determine coordinates of the electron not disturbing its further motion. The data obtained evidence that the similarity between the behavior of electrons and waves is somewhat exaggerated. The aim of the present research is to estimate the extent of this exaggeration.

\section{Diffraction-reradiation of electromagnetic waves and non-wave mechanism of the electron scattering at a half-plane (at a single sharp straight edge)}

I surmised that the similarity of the scattering patterns of electrons and EMWs comes from the periodicity of the structure of scatterers used in experiments of G.P.Thomson and C.Davisson (1927) and other authors at later years.  In order to escape the artifacts due to repeating character of the scatterer's structure I determined to compare the diffraction pattern of EMWs with that of electrons in a most simple case of a unitary scatterer.

\begin{figure}[h]
  \begin{center}
  \includegraphics[scale=0.7]{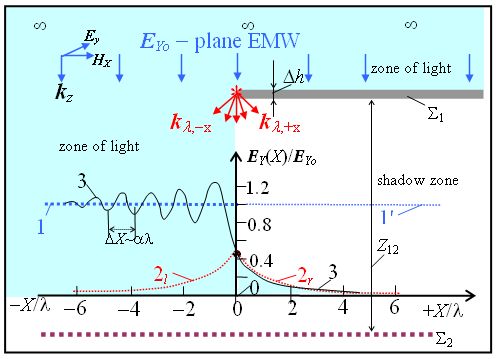}
  \caption{The diffraction (according to \cite{Born Wolf}) of normally incident electromagnetic wave (EMW) at the ideally conducting half-plane $\Sigma_1$ with the supposed thickness $\Delta h\rightarrow 0$. The distance from $\Sigma_1$ to the receptor screen  $\Sigma_2$ is equal  $Z_{12}=3\lambda$. The scattering edge is parallel to the $Y$-axis that is perpendicular to the plane of the figure. The area $+X/\lambda$ is the shadow zone beneath the half-plane, where there is no straight EMW, $|{\bf E}_{Y_o}|=0$, and there occurs only the right part ($2_r$) of the EMW diffracted (re-radiated) from the edge. The area $-X/\lambda$ is the non-screened by $\Sigma_1$ zone of light, where the interference sum ($3$) is composed of the incident EMW ($1$) and left part ($2_l$) of the  EMW diffracted from the edge.}\label{fig1}
\end{center}
\end{figure}

Such is the sharp edge of an opaque half-plane (see Fig.\ref{fig1}). The diffraction of EMWs is well studied both in theory and experiment. Its main features are represented on Fig.\ref{fig1} taken from \cite{Born Wolf}.

\begin{figure}[h]
  \begin{center}
 \includegraphics[scale=0.7]{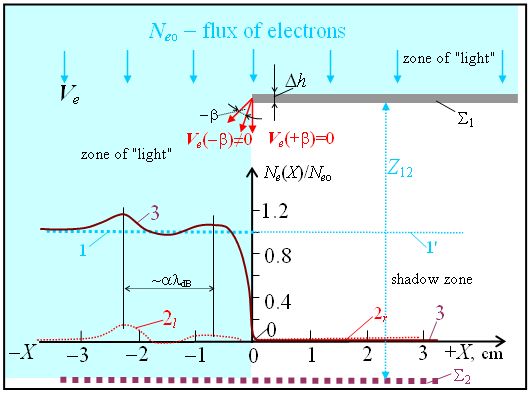}
  \caption{Ricochet scattering (according to \cite{Demjanov}) to the zone of "light" of the electron's flow by the sharp edge ($\Delta h\sim 4\lambda_{\texttt{dB}}$) of the opaque half-plane $\Sigma_1$. The distance $Z_{12}$ between $\Sigma_1$ and the visualization screen $\Sigma_2$ is about 1 m. The electron's distribution ($3$) shows the absence of the bending around the barrier to the shadow zone. The distance between peaks is proportional to the de Broglie wavelength $\lambda_{\texttt{dB}}$. The region $+X$ from the edge under the half-plane is the shadow zone of the screened straight flow; in this area there is no electrons scattered from the edge (the curve $2_r$). The region   $-X$ from the edge to the zone of "light" is formed by the primary beam ($1$) that is summed up  (curve 3) with the scattered ($2_l$) by the edge flow of electrons not interfering with it.
}\label{fig2}
\end{center}
\end{figure}

All distinctive features of scattering of EMWs (Fig.\ref{fig1}) and electrons (Fig.\ref{fig2}) are pristine in this experiment because of the uniqueness of the sharp scattering edge.  The scattering pattern of the primary beam of EMWs is determined by the symmetry of the angular spectrum of the the re-radiation of light by the linear verge: to the left from the edge, into the zone of "light", and to the right, into the shadow beneath the half-plane (Fig.\ref{fig1}). In the case of EMWs, both the experiment and theory indicate that the primary beam e.g. of $E_y$-polarized EMWs excites all Y-boundary of the half-plane $\Sigma_1$ making it similar to a luminescent filament. When exposing to EMW of the length $\lambda$  sharp edges of the thickness $\Delta h$ keep to shine up to $\Delta h \simeq 5\lambda$. In the end the light is re-radiated from the phase centers of the edge in the form of the angular spectrum of secondary EMWs that has the cylindrical symmetry with respect to this edge retaining the frequency  of the primary light's source (see also \cite{Morton}).

The right pattern (Fig.\ref{fig1}, hyperbola $2_r$) of re-radiated EMWs covers the shadow zone beneath the half-plane without interference since here there is only the re-radiation from the edge.  The left portion of re-radiated EMWs (hyperbola $2_l$) is added up with the principal primary beam of light, interferes with it and gives the curve (3). Its interference maxima and minima are repeating, in principle,  without limit over the interval $\Delta X$ that is multiple to the length $\lambda$ of EMW. While receding from the scattering edge, the amplitude of interference peaks decreases as $\sim 1/r$ until it will sink in the noise and become unobservable.

The exposing of the shadow zone (curve $2_r$) by the right part of the angular spectrum of EMW, that  radiates the edge,  is usually referred to as diffraction, or bending around the barrier by the primary flow. In the shadow zone the intensity of the re-radiated by the edge EMWs when receding from the edge decreases steadily according to the law $\sim 1/r$ as well (curve $2_r$). As a matter of fact this is not a bending around the obstacle by the primary beam but  quite a new radiation arising in the course of the two-stage wave process in which the secondary radiation from the edge does not  represent a direct continuation of the primary one.  The phase centers of this new source are located not at infinity (as those of the primary wave), but are firmly attached to the boundary of the half-plane. This process  drastically differs from processes of scattering of particles. I demonstrate this difference in experiments on scattering electrons \cite{Demjanov} (see below).

The experimental studies of the scattering of the electron beam at the sharp border of the opaque for electrons  half-plane showed that, firstly, there is no symmetrical angular dispersing of the fraction of the electron's flow (to the left and to the right) on the border. Electrons are scattered from the edge only to the zone of "light" (to the left in Fig.\ref{fig2}), and there is in effect no scattering à to the right, beneath the half-plane. Secondly, for all studied by me energies of electrons of the primary beam (0.2 eV through 100 eV) the most intensive flank peak (Fig.\ref{fig2}, curve $2_l$) produced by scattered electrons appeared to be uttermost. The distance to the peak is inversely proportional to the energy of the electrons. There is no scattering of electrons by the edge beyond this peak (i.e. at larger angles). Thus in scattering pattern of electrons there is no infinite spectrum of side peaks with the hyperbolically diminishing intensity peculiar to the diffraction of EMWs on the half-plane.

I saw it from that all sensors located at the distance farther of this peak read no addition to the primary flux  $N_{eo}$.  Thirdly, as more delicate measurements showed, in the pattern of electron scattering at the edge of the half-plane there occur weaker peaks (Fig.\ref{fig4}), but they tend to the center of the main maximum in the zone of "light". Their intensity also decreases in the reverse direction (i.e. as they approach to the spot under the scattering edge). The distance between peaks appears to be multiple of the de Broglie wave $\sim\alpha\lambda_{\texttt{dB}}$ where $\alpha(Z_{12})$ is the multiplicity coefficient of the experimental setup.

These observations can be interpreted in terms of the ricochet mechanism of scattering of electrons as particles with a complex inner structure. Electrons are scattered as if they have a set of concentric zones of elasticity   located from each other at the distance multiple to de Broglie wave length $\lambda_{\texttt{dB}}$.  It is difficult to explain the  diffraction pattern in the bounds of the conventional point model of the particle not using the notion of concentric elastic zones around the core of the particle.

\begin{figure}[h]
  \begin{center}
 \includegraphics[scale=0.7]{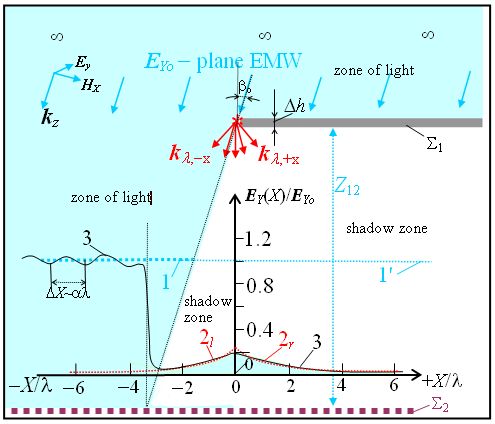}
  \caption{The diffraction (according to \cite{Demjanov}) of the oblique (at the angle $\beta_o\sim 35^\circ$) incident electromagnetic wave (EMW)(the wavelength $\lambda\sim 1 \textrm{cm}$) by the high conducting half-plane (copper). The  sharp scattering edge ($\Delta h\sim 0.05\lambda$) is parallel to the $Y$-axis that is perpendicular to the plane of the figure. The distance to the detecting screen is $Z_{12}=5\lambda$. The picture demonstrates that patterns $2_l$ and $2_r$ keep the attachment to the normal beneath the scattering edge at any $\beta_o$. The region $X/\lambda > 3.5$ is the shadow zone beneath the half-plane, where there is no direct EMW, $|{\bf E}_{Y_o}|=0$, and there occurs only the part ($2_l$) of the left and the whole right ($2_r$) distribution scattering by the edge EMW. The area $X/\lambda <  -3.5$ is the non-screened by $\Sigma_1$ zone of light, where the interference sum (3) is composed of the incident EMW ($1$) and the remote part ($2_l$) of diffracted from the edge EMW.
}\label{fig3}
\end{center}
\end{figure}

\begin{figure}[h]
  \begin{center}
  \includegraphics[scale=0.7]{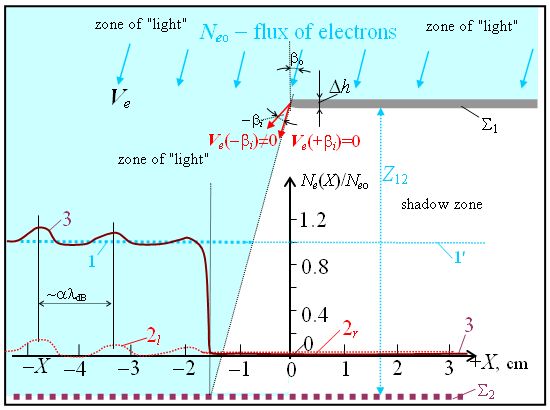}
  \caption{Ricochet scattering (according to \cite{Demjanov}) to the zone of "light" of the obliquely incident (at the angle $\beta_o\sim 1^\circ$) electron beam  by the sharp edge ($\Delta h=4\lambda_{\texttt{dB}}$) of the opaque half-plane. The distance from the scattering object to the registration screen is $Z_{12}\sim1 \textrm{m}$. The distribution testifies of the absence of the bending by electrons around the barrier to the shadow zone (there is registered the zeroth level $2_r$ of the flux). The area $X>-1.5$ cm is the shadow zone beneath the half-plane, where there is no electrons. The area $X<-1.5$ cm is the  zone of "light", where the primary (1) and scattered ($2_l$) flows of electrons  are summed up (to the curve 3).}\label{fig4}
\end{center}
\end{figure}

The wave nature of the diffraction of light in comparison with the ricochet scattering of electrons most distinctly manifests itself in the slanted (at some angle $\beta_o\neq 0$) incidence of light and electrons of the half-plane (Fig.\ref{fig3} and Fig.\ref{fig4}). We see from Fig.\ref{fig3} that the angle of bifurcation of the primary flow of light at the straight edge of the half-plane are firmly attached to phase centers of the re-emission of light by the edge, and this attachment is not changed for all $\beta_o\ne 0$.

A different picture is observed for electrons. Fig.\ref{fig4} implies that when $\beta_0\ne 0$ electrons, as before, bounce off only to the left and there occur no scattering to the right. Insofar as there is no re-emission of the electrons by the edge the scattering pattern is not linked to the normal dropped from the edge.

In the oblique incidence of EMWs and electrons there is observed another two phenomena. In the case of EMWs the increase of the incidence angle $\beta_o$ decreases the number of observed peaks in the zone of light due to widening the area of the non-interfering secondary radiation under the half-plane (Fig.\ref{fig3}). On the contrary, in the case of electrons the increase of the incidence angle increases the number of diffraction peaks observed in the zone of light (Fig.\ref{fig4}), although the shadow zone is widening as well. Such behavior of electrons bears in itself the features that are specific for the ricochet scattering of multi-shell micro-objects with the elastic module of concentric shells decreasing as following along the radius from the core of the electron to periphery.

\section{Comparing EMWs and electrons behavior in scattering at wave-equivalent obstacles}

\begin{figure}[h]
  \begin{center}
  \includegraphics[scale=0.7]{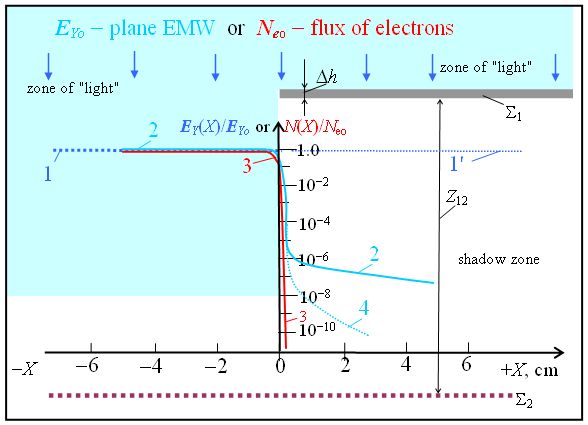}
  \caption{Comparing distributions in the shadow zone for scattering of electromagnetic waves (curve 2) and electrons (curve 3) at wave-equivalent edges of the half-plane. For EMW: $\Delta h=3\lambda$ and $Z_{12}=3000\lambda$. For electrons: $\Delta h=3\lambda_{\texttt{dB}}$ and $Z_{12}=3000\lambda_{\texttt{dB}}$. (1) is the distribution in primary flows without scattering half-plane. The curve 4 is the same as the curve 2, but with isolating the scattering edge by the EMW-absorbing stuff.}\label{fig5}
\end{center}
\end{figure}

In order to ascertain if electrons really behave as waves in scattering on sharp edges I performed the experiment on diffraction of EMWs (Fig.\ref{fig1}) with the configuration of scatterer wave-homothetic to that in scattering of electrons (Fig.\ref{fig2}). The distribution shown in Fig.\ref{fig1} was obtained with the registration diffracted EMWs in the near  zone of scattering, $Z_{12}=3\lambda$. The curve (2) in Fig.\ref{fig5} shows the distribution of EMW intensity in the far zone of scattering, $Z_{12}=3000\lambda$. Because of the latter the raised distribution of scattered EMWs in the zone of light actually degrade to the thickness of horizontal line (2). With the logarithmic scale used in Fig.\ref{fig5} the distribution of scattered electrons (curve 3) also pulls to the thickness of the horizontal line.

Fig.\ref{fig5} brings together the patterns for scattering EMWs and electrons with homothetic configurations of scatterer. It shows that with wave-reduced roundness radii of the edge (the thickness of the half-planes $\Delta h_{\textrm{EMW}}\simeq 3\lambda$ and $\Delta h_{\textrm{e}}\simeq 3\lambda_{\textrm{dB}}$) the scattering of electrons in the shadow zone is practically absent, whereas the re-radiation diffraction of EMWs is prominent.

Because of the multiple re-radiation of EMWs the special steps were taken to shield off other boundaries and elements of the (in practise finite) experimental setup. Otherwise the recording of the curve (2) would be impossible. In the case of electrons, which behave as microscopic bullets, these "safety precautions" were needless. The efficacy of the shielding is illustrated by curve (4) that was obtained via irradiation by EMWs the edge clothed by the EMW-absorbing stuff.

Scattering pattern shown in Fig.\ref{fig5} attest quite different mechanisms of scattering EMWs, on one side, and electrons, on the other side, at wave-equivalent objects.

\section{Diffraction of electromagnetic waves and non-wave mechanism of scattering of electrons by a slit (two sharp straight edges)}

\begin{figure}[h]
  \begin{center}
  \includegraphics[scale=0.7]{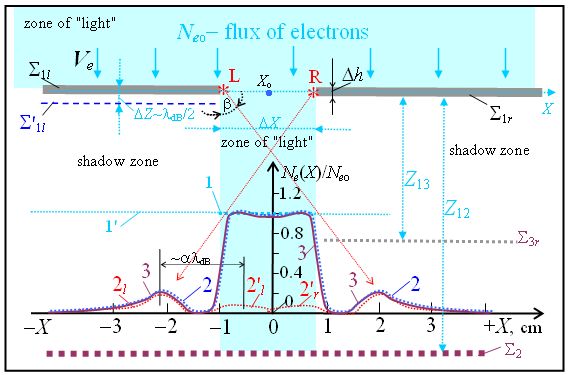}
  \caption{Ricochet scattering (according to \cite{Demjanov}) of the normally incident beam of electrons at the opaque plane ($\Delta h=5\lambda_{\texttt{dB}}$) with the slit of the width $\Delta X\sim 3\lambda_{\texttt{dB}}$ ($\lambda_{\texttt{dB}}$ is the length of the de Broglie wave). ($1$) is the distribution of electrons in the primary beam. ($2_r$) and  ($2'_r$) are right lobes of the scattering; they are formed solely by the left edge (L) of the half-plane $\Sigma_{1l}$. ($2_l$) and  ($2'_l$) are left lobes of the scattering; they are formed solely by the right edge (R) of the half-plane $\Sigma_{1r}$. (2)  is the "diffraction" pattern of electron's scattering on the slit formed by the half-planes $\Sigma_{1l}$ and $\Sigma_{1r}$ (with   $\Delta Z=0$). (3) is the "diffraction" pattern of the electron's  scattering on the slit formed by the half-plane $\Sigma'_{1l} $ shifted along the $Z$-axis by $\Delta Z\sim\lambda_{\texttt{dB}}/2$  and  $\Sigma_{1r}$ fixed (pattern 3 remains almost unchanged in comparison with 2). $\Sigma_{3r}$ is the moveable half-plane located at such a distance $Z_{13}$ from $\Sigma_{1r}$  that it hinders the flow of electrons scattered from the left edge (L) of the slit and suppresses the side lobe ($2_r$), but the zone of "light" is not screened.
}\label{fig6}
\end{center}
\end{figure}

The scattering pattern of electrons by a slit has a single peak on each side from the central maximum. As in contrast to the scattering of EMWs by a slit there is no any signs of "interference" beyond these peaks \cite{Demjanov}. In this experiment (Fig.\ref{fig6}) I realized for the first time that the left lobe of scattering pattern is concerned with the right edge of the slit, and the right one was formed due to ricochet from the left edge of the slit. This conjecture was then be verified by me in pure form by experiments on a unique boundary of the half-plane above described (Fig.\ref{fig2} and Fig.\ref{fig4}). In such a way the ricochet mechanism of scattering of electron by a slit was clarified.

There is another experimental proof that electrons scatter from edges of the slit by ricochet. It is simple (Fig.\ref{fig6}). I introduced a third half-plane $\Sigma_{3r}$ located at such distance $Z_{13}$ from the half-plane $\Sigma_{1r}$ that it intercepts the flow of electrons rebounded from the left ($\textrm{L}$) edge of the slit not affecting the zone of "light". Then the side lobe ($2_r$) disappeared, the left peak ($2_l$) staying as before. Similarly the left peak will be vanishing if a half-plane $\Sigma_{3l}$ be placed on the left at the same distance $Z_{13}$ from the half-plane $\Sigma_{1l}$. Below I will describe yet another means to prove validity of the ricochet mechanism for scattering of electrons at the pair of edges of the slit.

\begin{figure}[h]
  \begin{center}
  \includegraphics[scale=0.7]{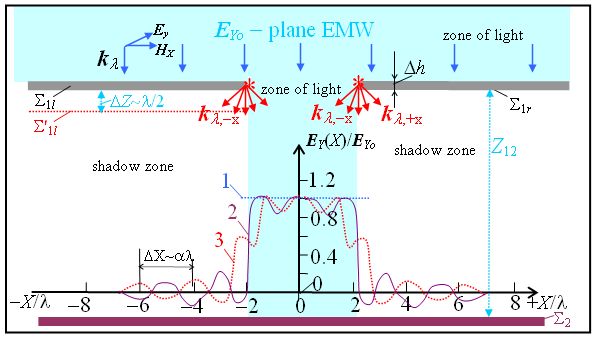}
  \caption{Diffraction (according to \cite{Demjanov}) of the normal incident flow of the electromagnetic wave (EMW) at the opaque half-plane (copper) with the slit of the width $\sim 2\lambda$ whose two scattering edges are parallel to the axis Y that is perpendicular to the plane of the figure; $\Delta h = 0.05\lambda$, $Z_{12}=5\lambda$. (1) is the amplitude of the incident wave (in the zone of light). (2) is the diffraction pattern for scattering of EMWs on the slit formed by the half-planes $\Sigma_{1l}$ and $\Sigma_{1r}$ which lie in the plane $\Sigma_1$. (3) is the diffraction pattern for scattering of EMWs on the slit formed by the half-planes $\Sigma'_{1l}$ and $\Sigma_{1r}$. $\Sigma'_{1l}$ is shifted along the $Z$-axis by $\Delta Z\sim\lambda/2$. The pattern (3) is almost the negative of the pattern (2).
}\label{fig7}
\end{center}
\end{figure}

But before I shall draw attention to that how the elementary complication of scattering system from one border of the half-plane (Figs.\ref{fig1}-\ref{fig4}) to two edges of the slit (Fig.\ref{fig6}) degrades the dissimilarity obtained on a one edge almost to identity of the scattering patterns of EMWs and electrons. Indeed when we deal with the scattering of EMWs at a single edge of the half-plane (Fig.\ref{fig1} and Fig.\ref{fig3}) the scattering pattern differs radically from that for scattering of electrons at the wave-equivalent single edge (Fig.\ref{fig2} and Fig.\ref{fig4}). While the scattering of electrons on two edges of the slit (Fig.\ref{fig6}) appears to be almost "twin" of diffraction pattern of EMWs on a wave-equivalent slit (Fig.\ref{fig7}) but for the absence of the infinite series of weak interference fringes in the case of electrons. So that, starting from the two edges of the narrow slit,  many revealed by me  peculiar distinctive features of the scattering of electrons at a single edge  (at the half-plane, Fig.\ref{fig2} and Fig.\ref{fig4}) are noticeably degrade  for two and more edges (as is seen in curve 3 of Fig.\ref{fig6}).

The differences in patterns of Fig.\ref{fig6} and Fig.\ref{fig7} could be noticed, in principle, if it were not for rapid recession of remote interference peaks for EMWs that makes them imperceptible in the noise. When there are many scattering elements as e.g. in crystal structures used in experiments of G.P.Thomson and C.Davisson (1927), the multitude of diffraction peaks of long-range order in EMW-patterns are overlapped resembling the raise of the noise. Against this background "noise" there remains noticeable  only the first pair of the most intensive side diffraction lobes  (Fig.\ref{fig7}), making the picture of the true diffraction of EMWs (curve 2 in Fig.\ref{fig7}) almost indistinguishable from the pattern of the ricochet scattering of electrons by the couple of edges (curve 3 in Fig.\ref{fig6}).

Fundamental differences of the scattering patterns of EMWs on narrow slit (Fig.\ref{fig7}) and electrons on wave-equivalent slit (Fig.\ref{fig6}) supplement results of experiments with small $Z$-shifts of one scattering edge with respect to its another fixed edge. These shifts $\Delta Z$ are shown in Fig.\ref{fig6} and Fig.\ref{fig7}.

For EMWs  the shift is of the order $\Delta Z\sim\lambda/2$, and in the case of electrons $\Delta Z\sim\lambda_{\texttt{dB}}/2$. The comparison of the curves (2) and (3) in Fig.\ref{fig7} for EMWs shows, that the shift of the left edge of the slit by $\Delta Z\sim\lambda/2$ relative to the right edge turns the curve (2) into the curve (3) as though converting negative to positive. For electrons the similar displacement of the left edge of the slit (Fig.\ref{fig6}) does not affect nor the main part of the pattern (3) (in the zone of "light") neither its side lobes ($2_l$) and ($2_r$). These experiments give convincing evidence of non-interference (non-wave) mechanism governing the formation of patterns for  scattering of electrons at sharp edges of narrow slits and, conversely, on patently wave-interference mechanism of the diffraction of EMWs at wave-equivalent narrow slits.

\section{The ricochet mechanism of scattering and the elastic-shells spatial structure of the electron}

So, in the diffraction of EMWs  on a slit there formed in the zone of light the main maximum resembling that formed in the scattering of electrons. To the left and right from the main maximum there are side peaks going with the interval $\sim\lambda$. The intensity of this peaks decreases as they recede from the center $X_0$ (Fig.\ref{fig7}). The phase centers of this interference sources lie at the left and right edges of the slit. Therefrom all the pattern of the diffraction-interference of EMWs depends greatly on the relative location of the slit's edges referring to the wave's length.  The absolutely different relationship is observed in the scattering of electrons at a wave-equivalent slit.

\begin{figure}[h]
  \begin{center}
  \includegraphics[scale=0.7]{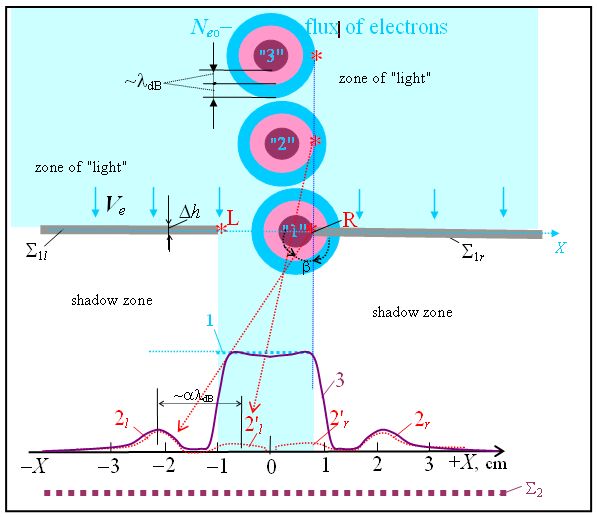}
  \caption{The model (according to \cite{Demjanov}) of the ricochet scattering of the normally incident beam of electrons by the right sharp edge ($\Delta h=5\lambda_{\texttt{dB}}$) of the narrow slit in the opaque screen $\Sigma_{1r}$. When the electron touches the slit's edge by its hard core it rebounds contributing to the remote lobe $2_l$. When the electron touches the edge by the soft shell "2" it rebounds on a less angle contributing to the lobe $2'_l$ which is nearer to the center of the slit. Electrons "3" pass by almost without deviation.
}\label{fig8}
\end{center}
\end{figure}

In the case of electrons the decreasing of the amplitudes of the left and right series of peaks is reversed, going from periphery to the central maximum (as is shown in Fig.\ref{fig2} and \ref{fig4}). So, there is observed only a couple of peaks (Fig.\ref{fig6}). All subsequent peaks are merged unnoticed in the main central peak of the zone of "light". I established that the right lobe (Fig.\ref{fig6}) is formed by the scattering of electrons from the left edge of the slit, and the left lobe is formed by the right edge. I found here that electron's scattering pattern does not depend on the shift of the left half-plane of the slit by $\Delta Z\sim\lambda_{\texttt{dB}}/2$ (see Fig.\ref{fig6}), the right half-plane being stationary (or vice versa). Such the behavior is not peculiar to EMWs (see Fig.\ref{fig7}). In this experiment the electron conducts itself not as a re-radiated by the edge wave and not as "probability amplitude  wave". It behaves as deterministic non-local micro-object with a spatial density distribution. From the experimental observations of the angles of preferred scattering we may infer that this density distribution has the shape of concentric shells of different hardness, the interval between the shells being of the order of de Broglie length (Fig.\ref{fig8}).

The scattering relief that we observe on the screen $\Sigma_2$ is predetermined by the respective structure of elastic shells inherent to the electron. The remotest peaks from the left and right arise due to the hard elastic interactions  of the electron's core "1" (Fig.\ref{fig8}) with the opposite slit's edge so that electrons rebound at the maximal angle of scattering. The experiment shows that these peaks are largest. The smaller peaks are formed due to the soft elastic shell (the electron "2" in Fig.\ref{fig8}) rebounding electrons at a less scattering angle. So they draw closer to the center. Electrons in the position "3" and partly in "2" are merged with the main maximum in the zone of "light" of the pattern.
\pagebreak
\enlargethispage{50pt}

\section{Interference of electrons on two closely spaced slits (four sharp edges)}

Diffraction on a single edge showed that the leading mechanism of scattering of electrons is the ricochet. It is not in the least like the diffraction of EMWs in this configuration of scatterer. Making the scatterer more complex, two edges (one slit), brought to that scattering patterns of electrons and EMWs acquired some common features. Diffraction and interference of electrons and EMWs on two slits (four edges) made this  similarity greater. However there are remained here some interesting differences that deserve special attention.

\begin{figure}[h]
   \begin{center}
  \includegraphics[scale=0.6]{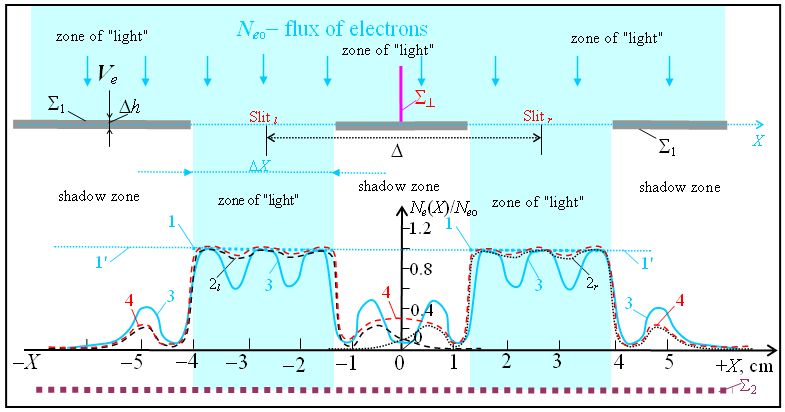}
  \caption{The  scattering (according to \cite{Demjanov}) of the normally incident flow of electrons by the opaque plane $\Sigma_1$ ($\Delta h\sim 3\lambda_{\texttt{dB}}) $with two slits (i.e. with four edges); the slit's width is $\Delta X\sim 3\lambda_{\texttt{dB}}$; the distance between slits $\Delta \sim 5\lambda_{\texttt{dB}}$. (1) is the intensity of electron's flow in the zone of "light"; (1') is the same in the absence of the screen $\Sigma_1$. ($2_l$) is the distribution (on $\Sigma_2$) of the electrons scattered from the left slit with the closed right slit. ($2_r$) is the distribution (on $\Sigma_2$) of electrons scattered from the right slit with the closed left slit. (3) is the distribution of electrons scattered by two open slits without the partition wall $\Sigma_\perp$. (4)  is the distribution (on $\Sigma_2$) of the electrons scattered by two open slits with the opaque for EMWs screen $\Sigma_\perp$ (it corresponds to the sum of partial distributions $2_l$ and $2_r$).}\label{fig9}
\end{center}
\end{figure}

On electron beams with the energy $\sim0.3\div1.0$ eV there was obtained the well-known effect \cite{Born}: the scattering pattern of two open slits (curve 3 in Fig.\ref{fig9}) differs from the sum (curve 4) of partial distributions obtained for each slit separately (curves $2_l$ and $2_r$ in Fig.\ref{fig9}).

\begin{figure}[h]
    \begin{center}
  \includegraphics[scale=0.6]{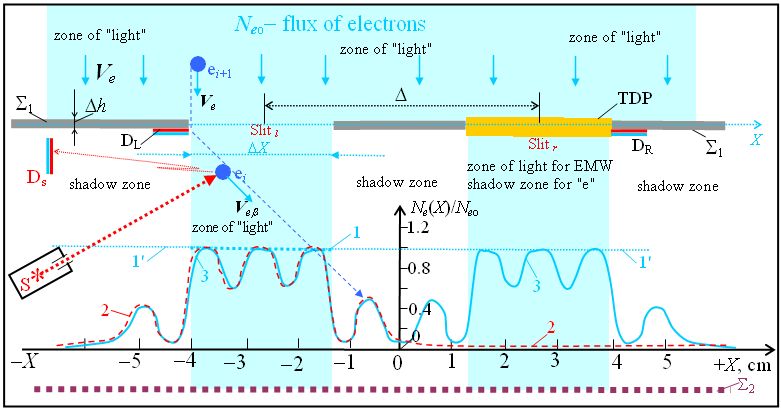}
  \caption{The scattering (according to \cite{Demjanov}) of the normally incident beam of electrons by the opaque plane ($\Delta h\approx 3\lambda_{\texttt{dB}}$) with two narrow slits of width $\Delta X\approx 3\lambda_{\texttt{dB}}$ spaced from each other by $\Delta\approx 5\lambda_{\texttt{dB}}$. (1) is the intensity of the electron's in the zone of "light"; (1') is the same in the absence of $\Sigma_1$. (2) is the distribution of electrons scattered by the left open slit and the right slit covered by the transparent dielectric plate (TDP). (3) is  the distribution (on $\Sigma_2$) of electrons scattered by two open slits. S is a lamp throwing light on electrons near the slit. D$_\textrm{s}$ detects the light scattered by an electron treated with the lamp S. D$_\textrm{L}$ and D$_\textrm{R}$ are semiconductor sensors of electrons arranged near left and right slits, respectively.}\label{fig10}
\end{center}
\end{figure}

Usually one tries to detect what slit has the electron passed through treating electrons near the slit with the side source $(\textrm{S})$ of light. However, illuminating electrons makes them stochastic (the scheme is shown on Fig.\ref{fig10}), and the interference disappears. The total pattern (curve 3 in Fig.\ref{fig9}) turns into the curve (4) that corresponds to the arithmetic sum of partial patterns ($2_l$ and $2_r$).

The same effects, destroying of the interference pattern, is attained if we will mount the opaque for EMWs copper partition wall $\Sigma_\perp$ (of the height not less than 1 mm) collinear to the flux of electrons. In the case of EMWs the  partition wall $\Sigma_\perp$ of the respective height does not actually influence the interference pattern. This experiment implicates that the electron interferes on the screen ($\Sigma_2$) with its EMW-component passing from the adjacent slit \cite{Demjanov}, and thus there is no need to adopt the generally accepted idea of a simultaneous passing of one and the same electron through two slits of the diffraction grating (Fig.\ref{fig9}) \cite{Tonomura}.
\pagebreak

\enlargethispage{150pt}
\pagebreak

In order to clarify this question the most difficult experiments were performed (see Fig.\ref{fig10}). I screened the right slit by the shield (really inserted in the slit a piece of dielectric) that entirely blocks up the passing of electrons through it. The shield (TDP) was made of the material transparent to light (UV rays). The scattering pattern under the former right slit got vanished. Yet the scattering pattern from electrons passed through the left slit (the left portion the curve 2 on Fig.\ref{fig10}) has remained almost the same as it was when both slits were open and without the partition wall $\Sigma_\perp$. When I screened the left slit by the same shield leaving the right slit open I found that the scattering pattern under the former left slit got vanished. Yet the scattering pattern from electrons passed through the right slit  have remained almost the same as it was when both slits were open. The scattering pattern of the configuration with two slits open (curve 3 in Fig.\ref{fig10}) appeared to be the the sum of partial scattering patterns just described \cite{Demjanov}.

\section{Detecting what slit did the electron passed through}

Passing of the electron through the slit always excites in its edges electric currents that can be basically measured. For example there may be mounted on the edge a micro-loop connected with an amplifier (that is difficult to put into practise \cite{Demjanov, Demjanov1}).  When taking dielectric, metal or semiconductor not specially adapted as detector these currents remain unnoticed.  Using as a scatterer the semiconductor with p-n junction I have succeeded in detecting the indications of the passing nearby electron. Was electron detected or not does not affect the scattering pattern.

It appeared that the passing through a slit electron excites another slit as well. For example, when I measured the current induced by the electron passed through the left slit ($\textrm{Slit}_l$, Fig.\ref{fig10}), there was simultaneously observed a weaker current in the diode sensor mounted at the right slit. In this event the scattering pattern was not affected by amplifiers linked with the sensors (cf. \cite{Mittelstaedt}).

\section{The influence of the temperature of scatterer on the scattering patterns of electrons and electromagnetic waves}

When I treated electrons near slits with the extraneous light according to the scheme of Fig.\ref{fig10} I have discovered another demonstration of the difference in the mechanisms of diffraction of electrons and EMWs on the slit. It is concerned with the warming-up of the edges of the slit due to the agency of illumination lamps (S in Fig.\ref{fig10}).

\begin{figure}[h]
  \begin{center}
  \includegraphics[scale=0.5]{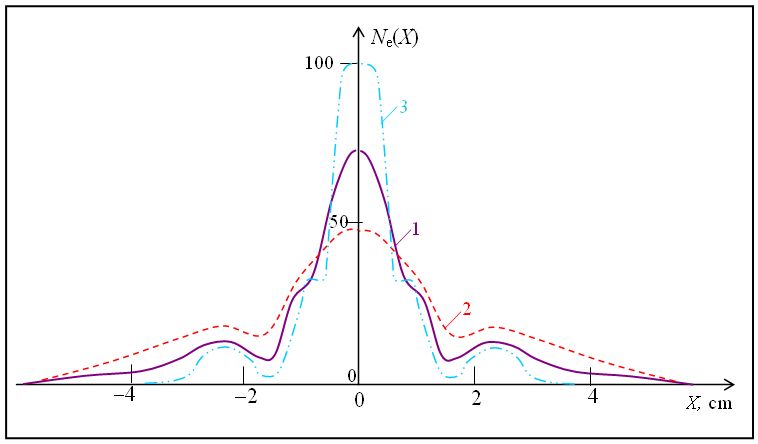}
  \caption{The distribution of electrons (according to \cite{Demjanov}) of the energy 0.6 eV scattered at the slit $\Delta X\sim 10^{-6}$ cm in the opaque screen  $\Sigma_1$. Curves 1 and 3 are taken at three temperatures of the screen $\Sigma_1$ in the location of slit according to Fig.\ref{fig6}: 1 $-$  $297^\circ$K;   2 $-$  $600^\circ$K; 3  $-$ $100^\circ$K. The distance $Z_{12}$ between the slit in $\Sigma_1 $ and the registration screen $\Sigma_2$ is about 1 m (see Fig.\ref{fig6}).}\label{fig11}
\end{center}
\end{figure}

The heat motion of atoms of the slit's edges appeared to blur heavily the scattering pattern of electrons (Fig.\ref{fig11}). The temperature variations of the electron patterns are much greater than the change of the geometrical proportions (dimensions) of the diffraction slit expected from its coefficient of thermal expansion.

Similar experiments on EMWs with heating of slits showed that diffraction patterns of EMWs were not virtually dependent on the temperature of slit's edges.

In these experiments electrons refuse to behave as waves. While EMWs do not conduct themselves as particles.

\section{Discussion and conclusion}

In the series of experiments above described I disproved the generally accepted since 1930s opinion that it is allegedly impossible  without destroying interference pattern to observe the place, say, a slit, where the electron passed. Performing the research with the exposing the slits to various sources (S) of light (by the scheme shown in Fig.\ref{fig10}), I confirmed that the strongly perturbing illumination is indeed unsuitable for observing electrons without demolition their interference patterns. Whereas using the semiconductor sensors fixed in the slit's edges enabled me to observe electron's fly not destroying the interference pattern. Since it does not matter for the electron if currents induced by him in the scatterer are detected or not detected.   With this it became clear that the electron's scattering pattern does not depend on the material of the slit's edges, be it a conductor, semiconductor or dielectric.

Semiconductor sensors perceptible to nearby electron were so efficient that appeared to be capable to determine not only the slit where the electron passed through but even the proximity of the track to the left or right edge of each aperture. So, the observation technique developed by me enables to an observer to determine  the trajectory of the electron's trip between two interactions: the tail of the trajectory lies at the edge of the slit currently observed and the end in the point where the electron encounters the interference screen. This calls in question the generally accepted interpretation of the electron as the "wave of probability". I found that at least in experiments on scattering electrons at singular edges the electron behaves as a deterministic object. Anyway, if some of the future experiments will fail to determine the trajectory of the electron we should not be rash to declare the electron as indeterministic object having no trajectory as it happened in 1920 years after experiments on scattering electrons by slits and singularities of the crystal lattice. Rather we must wait patiently the next breakthrough in the experimental skill to detect subtle signs of determinacy or classical probability of corpuscules.

The remake and revision of crucial experiments in physics performed by me in 1969-73 years (which I managed to publish only recently \cite{Demjanov, Demjanov1}), showed the great difference in wave manifestations of EMWs and electrons in experiments on diffraction and interference at wave-equivalent barriers. The signs of similarity seemed between the scattering patterns of electrons and diffraction and interference patterns of EMWs being cleaned out of features brought in by periodicity of the scatterer's structure leave no doubts that in the case under discussion the electron is not a re-radiated wave. The main indications of the difference between the scattering of electrons and diffraction of EMWs on straight slits are as follows.
\begin{itemize}
  \item When encountering an obstacle the electron does not bend around it, as EMWS do, but bounce off it as is appropriate to classical elastic macro-objects.
  \item  In the scattering pattern of electrons the left side peaks is formed by the ricochet from the right edge of the slit, and the right side peak is formed due to ricochet of electrons from the left edge of the slit, no interference of the electron's flows taking place.
  \item No occurrence of interference between two flows of electrons scattered by the edges of the slit becomes obvious from that the scattering pattern of electrons is independent on the displacement of one edge relative to other edge by $\Delta Z\sim\lambda_{\texttt{dB}}/2$. In the similar experiment with EMWs there happens the drastic change of diffraction pattern from positive peaks to negative ones. Thus the electron behaves not as a wave but as a particle.
  \item There is observed the reverse order of the intensities of side peaks in scattering of electrons. In the scattering of EMWs intensities of side peaks decrease as going away from the center peak with the spacing  equal to wavelength $\sim\lambda$, while in the scattering of electrons there are observed two peaks with the largest peak being the uttermost and the distance between peaks $\sim\lambda_{\texttt{dB}}$.
  \item The scattering pattern of electrons is strongly temperature-dependent in contrast to the scattering pattern of EMWs that is virtually not affected by temperature; the greater the temperature of the slit's edges the more is the dispersing of the scattering angles of electrons.
  \end{itemize}

The experiments described can be interpreted in terms of a composite spatial structure of the electron viewed as a system of concentric spherical shells of hardness. When the electron contacts with the edge of the slit by its tough core it rebounds from the edge forming the uttermost and most intensive peak in the flank of the scattering pattern. When the electron contacts with the edge by the soft shell, located the distance of the order  $\lambda_{\texttt{dB}}$ from the core, there are formed the close-by peaks (which are less intensive in experimental patterns).

When in double-slit experiments with electrons we screen one slit (e.g. the right slit as pictured in Fig.\ref{fig10}) by the dielectric shield that is transparent for EMWs (up to UV rays) but opaque for electrons there is conserved the interference pattern beneath the open slit. The other part of the pattern, that is under the screened slit, vanishes. Changing the shield to the left slit we keep the right part of the interference pattern. This experiment demonstrates the splitting of the electron on two near-by slits into two branches: the corporeal entity, proper the electron, (passed through one slit) and the electromagnetic wave part induced by it in the neighboring EMW-transparent slit.

Concluding it should be stated that the differences in mechanisms of scattering of true waves and electrons on tough barriers found in my experiments speak of the necessity to revisit and revise the problem at the modern level \cite{Fremont, Lindner} of the experimental skill.

\begin{acknowledgments}
The author is grateful to Dr V.P.Dmitriyev for valuable comments and assistance in the preparation of this paper.
\end{acknowledgments}


\begin{thebibliography}{99}
\bibitem {Thomson} G.P.Thomson, A.Reid (1927). Diffraction of cathode rays by a thin film, Nature 119: 890.
\bibitem {Davisson} C.Davisson, L.H.Germer (1927). Reflection of electrons by a crystal of nickel, Nature 119: 558–560.
\bibitem {Born Wolf} M.Born, E.Wolf, Principles of optics, 4th ed., Pergamon Press, London, New York, Paris, 1968.
\bibitem {Morton} N.Morton, Diffraction at a straight edge, Phys. Educ., v. 18, 207-208 (1983).
\bibitem {Demjanov} V.V.Demjanov, The aetherodynamic determinism of the Primodials, Ushakov State Maritime Academy, Novorossyisk,  2004, 568 p (in Russian).
\bibitem {Born} M.Born, Atomic physics, 7th ed., Blackie and Son Ltd, London, 1963.
\bibitem {Tonomura} A.Tonomura, J.Endo, T.Matsuda, T.Kawasaki, and H.Ezawa, Demonstration of single-electron buildup of an interference pattern, Am.J.Phys., v.57, No 2, 117-120 (1988).
\bibitem {Demjanov1} V.V.Demjanov, The aetherodynamic ins and outs of the relativity and quants, Ushakov State Maritime Academy, Novorossyisk,  2006, 448 p (in Russian).
\bibitem{Mittelstaedt} P.Mittelstaedt, A.Prieur and R. Schieder, Unsharp particle-wave duality in a photon split-beam experiment, Found.Phys., v.17, No 9, 891-903 (1987).
\bibitem {Fremont} F.Fremont, A.Hajaji, R. O. Barrachina, and J.-Y.Chesnel, A Young-type experiment using a single-electron source and an independent atomic-size two-center interferometer: the realization of a thought experiment, J. Physics: Conference Series, v. 88, 1-7 (2007).
\bibitem {Lindner} F.Lindner, M.G.Sch\"{a}tzel, H.Walther, A.Baltu\v{s}ka, E. Goulielmakis, F.Krausz, D.B.Milo\v{s}evi\'{c}, D.Bauer, W.Becker, and G.G.Paulus, Attosecond double-slit experiment, Phys.Rev.Lett., v.95, No 4, 040401 (2005).
\end{thebibliography}
\end{document}